\documentclass[aps,prl,showpacs,twocolumn,groupedaddress,superscriptaddress,nobibnotes]{revtex4-1}

\usepackage{blindtext}

\usepackage[colorlinks,
          linkcolor=black,
            citecolor=blue,
            urlcolor=blue
           ]{hyperref}
\usepackage{amsmath,amsthm,amssymb}
\usepackage{array}
\usepackage{subfigure}
\usepackage{graphicx}
\usepackage{fancyhdr}
\usepackage{color}
\usepackage{extarrows}
\usepackage{enumerate}
\usepackage{epstopdf}
\usepackage{bm}
\usepackage{natbib}
\usepackage{comment}
\usepackage{float}
\usepackage{ulem}
\usepackage[columnwise
]{lineno}

\begin{document}


\title{Particle-hole symmetry breaking in a spin-dimer system TlCuCl$_3$ observed at 100 T}

\author{X.-G.~Zhou}
\email{zhou@issp.u-tokyo.ac.jp}
\affiliation{Institute for Solid State Physics, University of Tokyo, Kashiwa, Chiba 277-8581, Japan}

\author{Yuan~Yao}
\email{smartyao@issp.u-tokyo.ac.jp}
\affiliation{Institute for Solid State Physics, University of Tokyo, Kashiwa, Chiba 277-8581, Japan}

\author{Y.~H.~Matsuda}
\email{ymatsuda@issp.u-tokyo.ac.jp}
\affiliation{Institute for Solid State Physics, University of Tokyo, Kashiwa, Chiba 277-8581, Japan}

\author{A.~Ikeda}
\affiliation{Institute for Solid State Physics, University of Tokyo, Kashiwa, Chiba 277-8581, Japan}

\author{A.~Matsuo}
\affiliation{Institute for Solid State Physics, University of Tokyo, Kashiwa, Chiba 277-8581, Japan}

\author{K.~Kindo}
\affiliation{Institute for Solid State Physics, University of Tokyo, Kashiwa, Chiba 277-8581, Japan}

\author{H.~Tanaka}
\affiliation{Department of Physics, Tokyo Institute of Technology, Tokyo 152-8551, Japan}

\begin{abstract}
  The entire magnetization process of TlCuCl$_3$ has been experimentally investigated up to 100 T employing the single-turn technique.
  The upper critical field $H_{c2}$ is observed to be 86.1 T at 2 K.
  A convex slope of the $M$-$H$ curve between the lower and upper critical fields ($H_{c1}$ and $H_{c2}$) is clearly observed, which indicates that a particle-hole symmetry is broken in TlCuCl$_3$.
By quantum Monte Carlo simulation and the bond-operator theory method, we find that the particle-hole symmetry breaking results from strong inter-dimer interactions.
\end{abstract}
\maketitle

Bose-Einstein condensation (BEC) is one of the most fascinating purely quantum-mechanical phenomena.
Early experiments had successfully realized the BEC state on dilute atomic gas system under nanokelvin temperatures with the laser cooling technique~\cite{BECATOM}.
The fact that nanokelvin temperatures is required makes experimental studies of BEC difficult to conduct and the number of the researches is limited.
On the other hand, based on a correspondence between a quantum antiferromagnet and a lattice Bose gas system~\cite{matsudabara1956},
the new research area on BEC  with spin ($s$) systems in which spins are strongly correlated to each other,  has been opened up in recent two decades.
It has been proposed that the field-induced ordering phase in TlCuCl$_3$ (a three-dimensional $s$=1/2 spin-gap dimer system) can be interpreted as a BEC phase of magnons~\cite{oshigawa2000,misguich2004,yamada2008,Sherman2010, Sherman2003},
and its magnetic field-temperature ($H$-$T$) phase boundary is claimed to follow the power-law dependence as,
\begin{eqnarray}
\label{power_law}
 g|H_{c1}(T)-H_g| \propto T^{\alpha},
\end{eqnarray}
where $g$ is the $g$-factor of the spin system, $H_{c1}$ is one of the critical magnetic fields of the BEC at a finite temperature $T$, $H_g$ is the critical magnetic field at zero kelvin, and the critical exponent $\alpha=1.5$ under Hartree-Fock-Popov (HFP) approximation.
It also provides a new experimental approach to study the BEC, i.e., magnons in three-dimensional spin-gap dimer system are studied as bosons in a lattice system.

In contrast to the ultra-cooling atomic gas, observations of critical behaviors around quantum critical points in the spin systems are accessible
because of their high degree of homogeneity in boson density and wide temperature window of the critical boundary ~\cite{Giamarchi2008}.
In a pure dimer system, the transition between singlet state and the $S_z=1$-triplet state under an external magnetic field drives a phase transition at the critical magnetic field $H_g\equiv\Delta/g\mu_\text{B}$ with $\Delta$ the spin gap.
The $H_g$ separates into two distinct critical points $H_{c1}$ and $H_{c2}$ in real spin-gap materials due to inter-dimer interactions; these two critical magnetic fields correspond to the begining and saturation of the magnetization, respectively.
The criticality in the vicinity of the $H_{c1}$ along the $H$-$T$ phase boundary has been extensively studied in several materials such as TlCuCl$_3$, Ba$_{3}$Cr$_{2}$O$_{8}$ and Sr$_{3}$Cr$_{2}$O$_{8}$, by means of magnetization, magnetostriction, neutron, and magnetocaloric effect (MCE) measurements~\cite{Johannsen2005,Tanaka2001,nohadani2004, kawashima2005,Aczel2009_1, Aczel2009_2}.
On the other hand, detailed experimental measurements at the high-field side near $H_{c2}$ are still lacking for materials that possess large $H_{c2}$ like TlCuCl$_3$.

In the theoretical analyses on $H_{c2}$ of BaCuSi$_2$O$_6$ ~\cite{Jaime2004}, the transformation from a spin Hamiltonian to an effective hard-core boson model was derived by projecting the original spin states to the low-energy singlet and the $S^{z} = 1$-triplet states.
Here, the singlet and $S^{z} = 1$-triplet states, respectively, correspond to the holes (particles) and particles (holes) of bosons at a low (high) magnetic field.
A particle-hole symmetry~\cite{mila1998} in this formulation is explicit in that the effective Hamiltonians at the low magnetic field and the high field are related by a particle-hole transformation on the bosonic creation operators and, thus, the magnetic-field dependence of the order parameters in phase diagrams has a symmetry with respect to the particle-hole invariant point $H=(H_{c1}+H_{c2})/{2}$.
For instance, such a symmetry is reflected by the mirror symmetry observed in the shape of the phase boundary of the BEC phase in BaCuSi$_2$O$_6$. Actually, the symmetric nature is verified in the bond operator theory~\cite{wang2006} by showing equality of the critical exponents in the vicinity of the two critical points, $H_{c1}$ and $H_{c2}$. Similar symmetric phase diagram was also obtained in Ba$_3$Cr$_2$O$_8$~\cite{Aczel2009_1}.
In the following discussion,
we will call such a \textit{phase-diagram} symmetry as ``particle-hole symmetry''.

It should be noted that the critical behaviors near the $H_{c2}$ in the system of which inter-dimer interactions are strong has never been well investigated. In such systems, the particle-hole symmetry of the BEC phase boundary is expected to be broken because the conditions of the hard-core boson picture are not  satisfied due to the effect of the hybridization between the singlet and $S_z=-1,0$-triplets.
According to the bond operator theory, the pressure effect (i.e. the changing of the inter-dimer interaction) on the phase boundary in the $H$-$T$ plane  is different at $H_{c1}$ and $H_{c2}$ \cite{Matsumoto:2004aa}, indicating that the symmetry can be modified by the inter-dimer interaction.
TlCuCl$_3$ is one of the most promising materials to study the symmetry of the phase boundary and discuss the particle-hole symmetry, because it possesses strong inter-dimer interactions.
$H_{c1}$ is 5.6~T~\cite{oosawa1999} and $H_{c2}$ is theoretically predicted~\cite{matsumoto2002} to be about 87~T; a strong contrast of these two critical points provides us a unique situation where the degrees of the quantum hybridization between the singlet and triplet states are different between two critical points. However, $H_{c2}$ has never been experimentally observed yet in the first-discovered magnon BEC compound, TlCuCl$_3$ because of technical difficulties to perform a 100-T class magnetic field experiment.

In this Letter, we study the magnetization ($M$) process of TlCuCl$_3$ up to 100 T and discuss the particle-hole symmetry breaking of the BEC phase. The particle-hole symmetry in the hard-core boson picture has been found to be broken through the quantum hybridization effect due to the strong inter-dimer interactions. $H_{c2}$ is experimentally determined to be $86.1$~T at $T\approx2$ K that is in good agreement with the prediction by the bond-operator theory~\cite{matsumoto2002}.
The magnetization curve ($M$-$H$ curve) has a monotonous convex downward shape, which manifests the difference between the critical exponents around $H_{c1}$ and $H_{c2}$ and implies the absence of the particle-hole symmetry in TlCuCl$_3$.
We have also precisely analyzed the $M$-$H$ curve with a numerical calculation based on Quantum Monte Carlo (QMC) method.
The simulation shows that the ratio of an inter-dimer interaction ($J_2$) and the intra-dimer interaction ($J$),  $J_2$/$J$ is about 0.35,
which is consistent with the results of the neutron-scattering experiment~\cite{matsumoto2002}.
Finally, we theoretically evaluate the critical exponents at near $H_{c1}$ and $H_{c2}$ and discuss on the microscopic origin of the breaking of the particle-hole symmetry using a combinatorial approach of the bond-operator theory and QMC calculation.

A single crystal of TlCuCl$_3$ \cite{oosawa1999} was used for the experiments. The single-turn coil (STC) technique was employed to generate a pulse magnetic field of up to 110 T. The magnetization measurements have been done with a pick-up coil and an induction voltage in the magnetization process is recorded as a function of time~\cite{Matsuda2013,takeyama2012}.
The magnetization curve is obtained by a numerical integration of the measured $dM/dt$ signal~\cite{Matsuda2013,takeyama2012}.
In the present work, we use a double-layer pick-up coil (the number of the turns of the coil [80 in total] is doubled from the standard number\cite{takeyama2012}) which improves the signal intensity. The magnetization signal $dM/dt$ is obtained by subtraction of the background signal from the sample signal. Here, the background signal is obtained by a measurement without the sample and the sample signal is a measured signal with the sample. The set of the two signals is obtained by two successive destructive-field measurements. This simple manner has advantages compared to the previous sample-position-exchange manner \cite{takeyama2012} in regard to  stability of the measurement position and effects from inhomogeneity of the magnetic field.
A liquid helium bath cryostat with the tail part made of plastic has been used \cite{takeyama2012};
the sample was immersed in liquid helium, and a measurement temperature of 2 K was reached by reducing the vapor pressure.

\begin{figure}[b]
    \includegraphics[width = 1.0\linewidth]{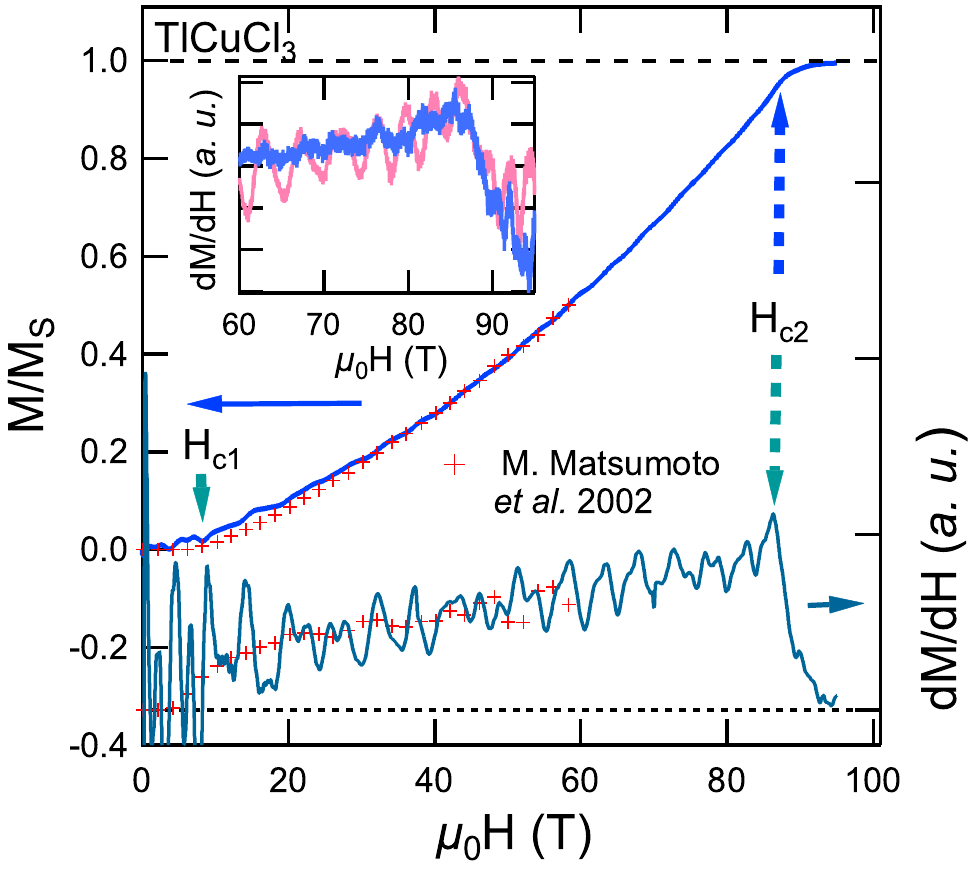}
     \caption{(Color online) The magnetization curve measured up to 95 T, as well as the $dM$/$dH$ data. Red marks represent the results of Ref.~\cite{matsumoto2002}. The inset shows the $dM/dH$ data (up-sweep and down-sweep) with initial temperature at 2 K.}
     \label{exp_M}
    \end{figure}

Fig.~\ref{exp_M} shows the magnetization process and the magnetic field dependence of the $dM/dH$ at 2 K;
The magnetization up to 60 T in Ref.~\cite{matsumoto2002} is also presented for the comparison of the $M$-$H$ curves. The agreement of them is excellent.
The $M$-$H$ curve is obtained from the average of three independent experiments using the both up-sweep and down-sweep processes; we do not use the data of magnetization below 70 T of the down-sweep for the average because the measurement in the field-decreasing process is rather imprecise due to the field inhomogeneity~\cite{Matsuda2013}.
We analyze the $H_{c1}$ around 6 T using the data previously obtained with a non-destructive pulse magnet \cite{matsumoto2002},
because, at the beginning of destructive ultrahigh magnetic field generation, a huge switching electromagnetic noise is inevitably generated for injection mega-ampere driving currents~\cite{Matsuda2013,takeyama2012}.
The magnetization is clearly measured in fields from 30 to 100~T, exhibiting the saturation at fields exceeding 90~T.

The critical behavior is observed in the vicinity of the saturation in the $dM/dH$ curve as shown in Fig.~\ref{exp_M}:
A clear peak labeled by $H_{c2}$ in $dM/dH$ curve corresponds to a kink of the $M$-$H$ curve observed at near $H_{c2}$.
The magnetization increases continuously  between $H_{c1}$ and $H_{c2}$, in which the $M$-$H$ curve shows  a convex slope.

Due to the fast sweep rate of the pulsed magnetic field ($\sim$ 8 $\mu$s), the sample undergoes a semi-adiabatic process.
The spin entropy change caused by a phase transition is transferred into the phonons and leads to the temperature change of the sample.
Therefore, it is necessary to consider the temperature change during the magnetization process. Because the MCE experiment up to 100~T is impossible to be conducted with current experimental techniques, we estimate the MCE effect using the MCE reported in another magnon BEC dimer spin system.
According to Ref.~\cite{Jaime2004}, the temperature change at 45 T near $H_{c2}$ in BaCuSi$_2$O$_6$ is around 0.03 K where the specific heat is 1.8~J/(mol~K) at 2.5 K. Then entropy change $\Delta$$S$=0.02~J/mol is estimated to be
transferred from the magnons to phonons around phase transition point.
Because the entropy change in the spins at $H_{c2}$ should reflect the way of the saturation of the magnetization where the BEC phase is terminated,
the entropy change should be of more or less a similar value as other spin systems which undergo magnon BEC with spin-1/2 dimers.
Therefore, the change of entropy observed in BaCuSi$_2$O$_6$ can be applied to the analysis on TlCuCl$_3$.
The specific heat in TlCuCl$_3$ contributed by phonons is  $C_\text{phon}$=4~J/mol~K at an initial temperature $T_\text{in}$=2~K   \cite{oosawa2001}, and thus the temperature rise from the initial temperature is estimated to be $\Delta$$S$$\cdot$$T_\text{in}$/$C_\text{phon}$=0.01~K  for TlCuCl$_3$ at near $H_{c2}$.
The insert of Fig.~\ref{exp_M} shows d$M$/d$H$ for both field up-sweep and down-sweep processes. The overlapping of both processes implies that the temperature variation is actually small.
Thus, critical points and critical exponents can be confirmed and analyzed at 2 K in the present work.\vspace{-0mm}

$H_{c2}$ is determined by the peak in the $dM$/$dH$ curve and found to be 86.1$\pm$0.5~T. It is very close to the predicted value obtained from the bond operator theory~\cite{matsumoto2002}. Furthermore, the continuous slope of the $M$-$H$ curve keeps its convex shape until $H_{c2}$. It indicates that the critical exponents can be potentially different between $H_{c1}$ and $H_{c2}$ because, otherwise, it is expected that $M$-$H$ curve at $H_{c1}$ and $H_{c2}$ should present the particle-hole symmetry behavior, i.e. the correspondences of the shapes at $H_{c1}$ and $H_{c2}$ are suppose to be linear versus linear, concave versus convex or convex versus concave.
    \begin{figure}[b]
    \includegraphics[width = 1.0\linewidth]{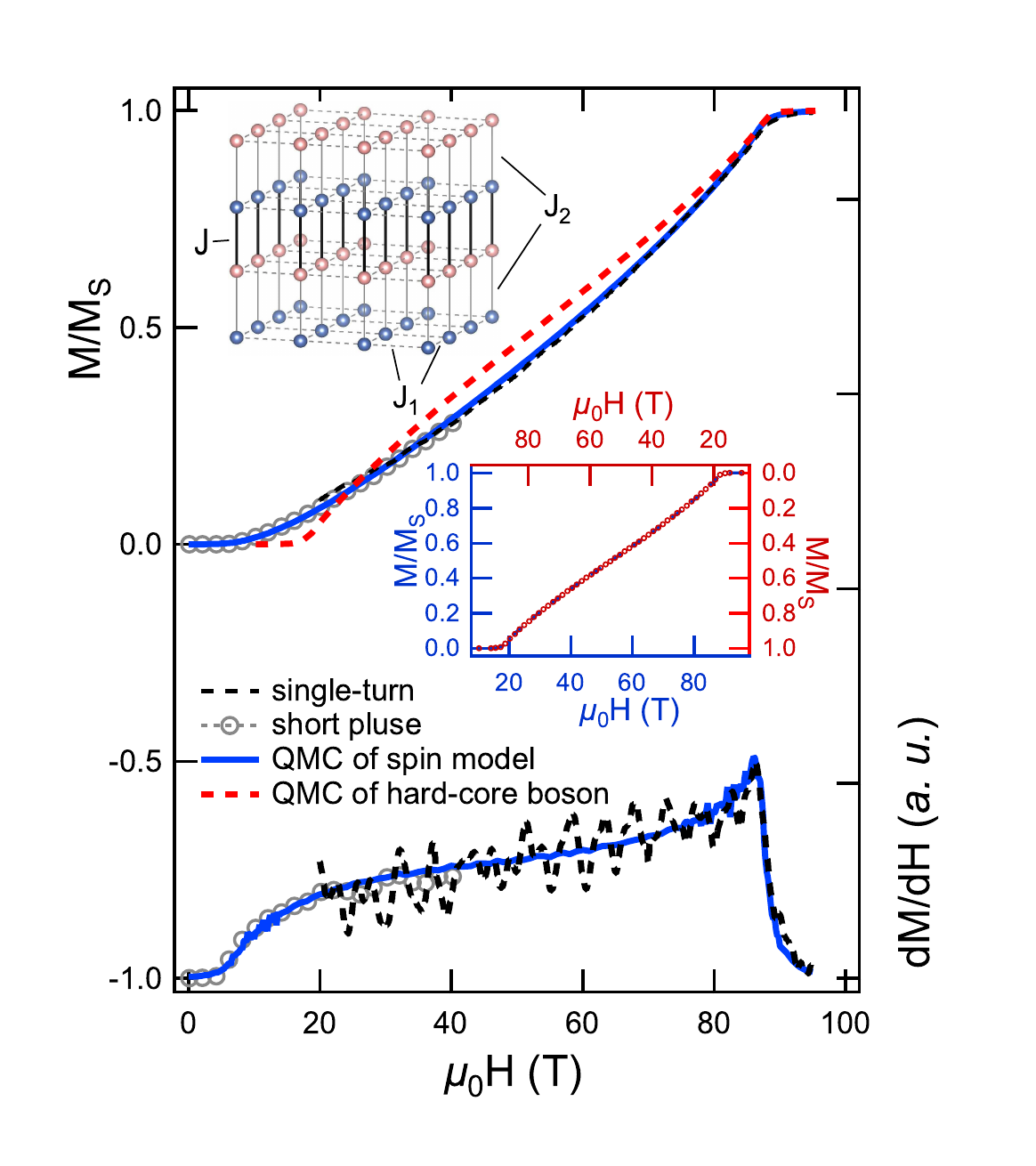}
     \caption{Comparison between the experimental magnetization curve and the QMC simulation results, the low field data (below 40 T) is from Ref.\cite{matsumoto2002}, the blue and red curves is obtained from Eq.~\ref{spin} and Eq.~\ref{Heff}, respectively. The upper inset shows the lattice of spin Hamiltonian. The red and blue ball represent the spin site 1 and 2 of a dimer in Eq.~(\ref{spin}), respectively. The middle inset shows the symmetry presentation in magnetization from Eq.~\ref{Heff}.}
     \label{comparison}
    \end{figure}
In order to interpret the observed magnetization process in more detail, we have tried to theoretically reproduce the $M$-$H$ curve with a rather simple spin model.
{The lattice Hamiltonains in previous study are complex and include several types of inter-dimer interactions~\cite{matsumoto2002,Matsumoto:2004aa}.}
Let us consider the following spin Hamiltonian to describe the universality classes around $H_{c1}$ and $H_{c2}$:
\begin{eqnarray}
\label{spin}
\mathcal{H}_\text{spin} &=&J\sum_{\mathbf{i}}\mathbf{S}_{\mathbf{i},1}\cdot \mathbf{S}_{%
\mathbf{i},2}+J_{1}\sum_{\mathbf{i}}\sum_{m}\sum_{n=x,y}\mathbf{S}_{\mathbf{%
i},m}\cdot \mathbf{S}_{\mathbf{i}+{\hat{e}}_{n},m}  \notag \\
&+&J_{2}\sum_{\mathbf{i}}\mathbf{S}_{\mathbf{i},2}\cdot \mathbf{%
S}_{\mathbf{i}+{\hat{z}},1}-g_{\parallel }\mu _{B}H\sum_{\mathbf{i},m}S_{\mathbf{%
i},m}^{z} \label{Hs},
\end{eqnarray}%
where $\mathbf{i}$ denotes the site of dimers in a cubic lattice, $m$ = $1$, $2$ shows the position of the two spins in one dimer, $\hat{e}_{n}$ = $\hat{e}_{x,y,z}$ represents the unit vector.
Here $J$ denotes the intra-dimer interaction, and $J_{1,2}$ are inter-dimer interactions in different directions.
We have set two different inter-dimer interactions $J_{1}$ and $J_{2}$, because it is necessary to fit and reproduce the experimental $M$-$H$ curve in Fig.~\ref{comparison} and their values will be shown later.
The assumed lattice for the $\mathcal{H}_\text{spin}$ is shown in the insert of Fig.~\ref{comparison}.
The QMC calculation is performed using a generalized directed loop algorithm in the stochastic series expansion representation~\cite{Wessel2005}, as implemented in the ALPS package~\cite{ALPS}.
The calculation is performed with 10$\times$10$\times$10 unit cells, in which each unit cell contains two spin sites.
Additionally, it has been reported that the structure of TlCuCl$_3$ is not magnetically frustrated, although the low-symmetry structure results in many triangle interaction terms in the Hamiltonian~\cite{Matsumoto:2004aa}.

In Fig.~\ref{comparison}, the magnetization curves simulated by the QMC calculation are shown with the experimental data.
The simulated $M$-$H$ and d$M$/d$H$ curves based on the Hamiltonian in Eq.~(\ref{Hs}) is represented with blue curve. 
The g-factor 2.23~\cite{oosawa1999} is used in this simulation.
The parameters determined in this calculation are 
$J$=5.318~meV, $J_1$=1.074~meV and $J_2$=1.86~meV.
In our reduced model, $J$ and $J_2$ agree with those of the original model of TlCuCl$_3$ in Ref.~\cite{matsumoto2002, Matsumoto:2004aa} according to their spin configurations. These values are also close to those reported in the previous work using the bond operator method~\cite{matsumoto2002}.
The simulated $M$-$H$ curve performed at $T$ = 3.5 K is found to agree very well with the experimental magnetization process.

The hard-core boson model constructed by only the $S_z$=1 triplet and singlet states is another way to analyze the process of magnetization.
Following Jaime \textit{et al.}~\cite{Jaime2004}, we translate the spin Hamiltonian (Eq.~\ref{spin}) to an effective Hamiltonian of the hard-core boson:
\begin{eqnarray}
\label{effective}
\mathcal{H}_\text{eff} &=& t_1 \sum_{\mathbf{i, \alpha}} (b^{\dagger}_{\mathbf{i}+{\hat e%
}_{n}}b^{\;}_{\mathbf{i}}+ b^{\dagger}_{\mathbf{i}} b^{\;}_{\mathbf{i}+{\hat e}%
_{n}}) + t_2 \sum_{\mathbf{i}} (b^{\dagger}_{\mathbf{i}+{\hat z}}b^{\;}_{%
\mathbf{i}}+ b^{\dagger}_{\mathbf{i}} b^{\;}_{\mathbf{i}+{\hat z}})  \notag \\
&+& V_1 \sum_{\mathbf{i,\alpha }} n_{\mathbf{i}} n_{\mathbf{i}+{\hat e}_{n}} +
V_2\sum_{\mathbf{i}} n_{\mathbf{i}} n_{\mathbf{i}+{\hat z}} + \mu \sum_{%
\mathbf{i}} n_{\mathbf{i}}  \label{Heff}
\end{eqnarray}
where the chemical potential is $\mu = J - g_{\parallel} \mu_B H$, and hopping terms are $t_1 = V_1 =
J_1/2$ and $t_2= V_2= J_2/4$. 
The simulation shares the same interaction parameters with the spin model.
Based on this reduced model,
we calculate the magnetization and show the results with the red dashed curve in Fig.~\ref{comparison}.
In contrast to the result of $\mathcal{H}_\text{spin}$, the magnetization derived from $\mathcal{H}_\text{eff}$ shows the central symmetry that relates the magnetizations around $H_{c1}$ and $H_{c2}$, which corresponds to the particle-hole symmetry shown in the middle insert of Fig.~\ref{comparison}.
However, in the following, we will see that such a symmetry is not held when the $J_2$-inter-dimer interactions in Eq.~(\ref{spin}) is significant.

As shown in Fig.~\ref{comparison}, the magnetization gradually increases with magnetic fields when the field approaches to and exceeds $H_{c1}$, while the magnetization drastically changes its slope at near the saturation field $H_{c2}$ in our experiment.
This behavior can be more clearly recognized in the d$M$/d$H$ curve; one
can see that rather sharp peak structure appears just below $H_{c2}$ and only a round shoulder structure is seen in the vicinity of $H_{c1}$.
This asymmetric character of the magnetization indicates a sizable breaking of the particle-hole symmetry.

    \begin{figure}[h]
    \includegraphics[width = 1.0\linewidth]{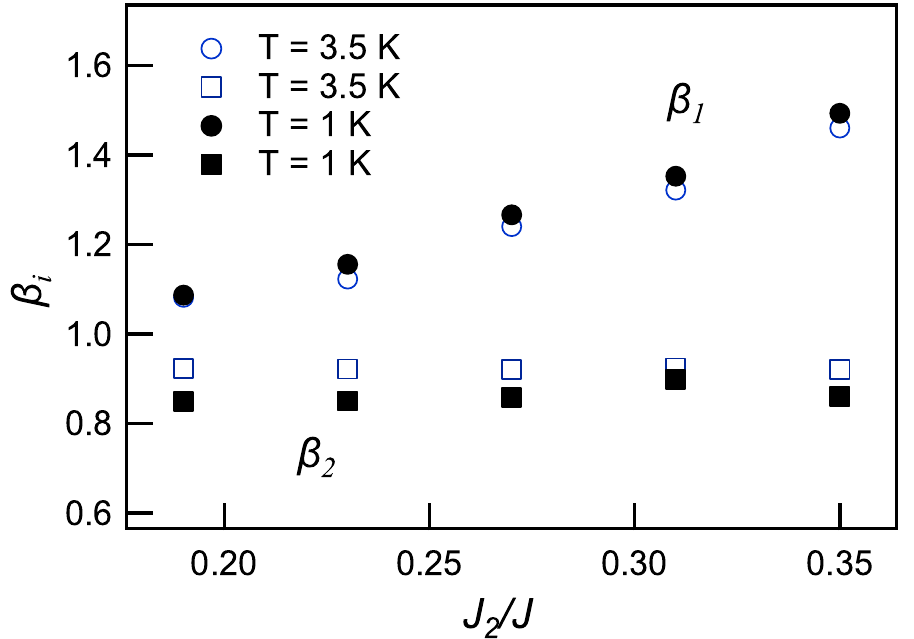}
    \caption{$\beta_{i}$ (i = 1, 2) as a function of $J_2/J$ with fixed $J_1/J=0.201$. The determination is performed with simulated $M$-$H$ curves at 3.5 K and 1 K. Circles and squares present $\beta_1$ and $\beta_2$, respectively.}\label{beta}
    \end{figure}

In order to clarify the asymmetric character, we utilize the following formula to perform the fitting of the magnetization.
 \begin{eqnarray}\label{2}
  |M-M_{ci}| \propto |H - H_{ci}|^{\beta_i}, \ \ (i = 1,2).
 \end{eqnarray}
The breaking of the particle-hole symmetry can be identified by comparing two critical exponents $\beta_{1,2}$ ~\footnote{Here, the critical exponents $\beta_{1,2}$ mean the the high-order term of variation, and considered to be ``effective'' critical exponent} around $H_{c{1,2}}$ in Eq.~(\ref{2}) since the particle-hole symmetry implies $\beta_1=\beta_2$.

By analyzing our experimental data with the QMC simulation, we can computationally determine the critical exponents with the simulated $M$-$H$ curve~\cite{hayashida2019}. We find that $\beta_2$ is robustly localized at 0.91. Here $\beta_1\sim 1.4$ is also obtained in the same analysis but using the experimental results obtained with a non-destructive magnetic field~\cite{matsumoto2002}.
The amount of the difference between $\beta_1$ and $\beta_2$ directly measures the degree of the asymmetry, and the difference has been found to depend on the degree of the inter-dimer coupling (itinerancy of the triplons).
We have conducted an analysis of the dependence of the exponents $\beta_{1,2}$ on the inter-dimer exchange interaction $J_2$.
In the procedure for determination of $\beta_{1,2}$, we define field windows $x_{1,2}={(H-H_{c{1,2}})}/{(H_{c2}-H_{c1})}$ which the data used should be within~\cite{nohadani2004}.
Accuracy of fitting process can be confirmed by stability of the obtained value when we change the field window.
The fittings are performed under $x_{1,2}$ = 0.1, 0.15 and 0.2.
All the obtained exponents are plotted as a function of $J_2$/$J$ in Fig.~\ref{beta}.
The range of $J_2/J$ is set from 0.18 to 0.35.
As shown in Fig.~\ref{beta}, $\beta_1$ gradually increases with $J_2$/$J$ while the value of $\beta_2$ is more robust.
The resulted $J_2$-variations of $\beta_1$ and $\beta_2$ show similar behavior at 3.5 K and 1 K, indicating that the qualitative dependence of the critical exponents on the effects of the inter-dimer interaction is not affected by changing temperature from 1 to 3.5~K.

The differences between $\beta_1$ and $\beta_2$ are reflected by the convex slope of the magnetization around $H_{c1}$ and $H_{c2}$ as in Fig.~\ref{comparison}; the initial magnetization process is suppressed compared to a linear magnetization while the slope increases when approaching the saturation, resulting in the larger $\beta_1$ than $\beta_2$.

In the bond operator theoretical formalism, the magnetism around $H_{c1}$ is contributed by the outset of lowest-triplet $|\uparrow\uparrow\rangle$, namely $S_z=1$-condensations driven by the vanishing spin gap.
In the pure-dimer limit $J\gg J_{1,2}$,
the other two higher-energy local excitations $(|\uparrow\downarrow\rangle+|\downarrow\uparrow\rangle)$ and $|\downarrow\downarrow\rangle$ can be neglected since they hardly affect the low-temperature physics.
However, a finite $J_2$ interaction can mediate a four-particle interaction that accumulates highest-triplet $|\downarrow\downarrow\rangle$ excitations in the presence of a partial lowest-triplet condensation and singlet condensation~\cite{Matsumoto:2004aa}.
Therefore, $|\downarrow\downarrow\rangle$ excitations suppress the magnetization from the $|\uparrow\uparrow\rangle$ condensations when $H\gtrsim H_{c1}$.
When the magnetic field increases beyond $H_{c1}$, such a suppression eventually becomes inefficient due to a large Zeeman gap to excite a $|\downarrow\downarrow\rangle$ quasiparticle and is neglibible near $H_{c2}$.
Therefore, we have the increasing deviation of $\beta_1$ from $\beta_2$ as shown in Fig.~\ref{beta}. The convex shape of the $M$--$H$ curve  at around $H_{c1}$ can be understood in the same way.
Moreover, it is found that $\beta_2$ is consistently insensitive to a changing of $J_2/J$.
In short, the deviation of $\beta_1$ from $\beta_2$ is attributed to the strength of $J_2$-inter-dimer interactions.
Therefore, it can be concluded that the particle-hole symmetry in the hard-core boson picture (Eq.~\ref{effective}) is broken in TlCuCl$_3$ due to the strong higher-order terms through the inter-dimer interactions.
Here, it should be noted that the similar inter-dimer effect is also obtained by increasing $J_1$ instead of $J_2$. 
The hybridization of  the four-particle states becomes relevant when $J_1$ or  $J_2$  (or both) is (are) sufficiently large.

Such a particle-hole symmetry breaking has been observed in another spin-${1}/{2}$ dimer system~\cite{Brambleby2017}. 
In that case, the origin of the asymmetry is claimed to be due to a zero-point quantum fluctuation characterized by the additional zero-point energy to quasi-particle excitations~\cite{kohama2011},  which should be signifiant in low dimensional systems.
However, such a quantum fluctuation in low dimensions is expected to be insignificant in TlCuCl$_3$ because the spin system is almost three-dimensional
~\footnote{Because the rate of magnetic susceptibility 
between $H_{c1}$ (5.6 T) and $H_{c2}$ (86.1 T), i.e. $\chi_1$/$\chi_2$ in TlCuCl$_3$ is as large as 5 (see Fig. 1) which is 2.5 times larger than that of NiCl$_2$-4C(NH$_2$)$_2$ Ref.~\cite{kohama2011}, the renormalized mass m$^*$ is expected to be reduced by a factor of (2.5)$^{3/2}$ $\sim$4 compared to the case of Ref.~\cite{kohama2011}. 
Therefore the m/m$^*$ in TlCuCl$_3$ should be as large as 3 $\times$ 4 $\sim$ 12. This seems to be too large for 3-dimensional spin system TlCuCl$_3$, and thus the quantum fluctuation may not be the origin of the asymmetry in TlCuCl$_3$.
}.

Furthermore, due to the insignificance of higher-energy triplet excitations around the saturation,
it can be expected that the $H$-$T$ phase boundary around $H_{c2}$ obeys a power law $g|H_s-H_{c2}(T)| \propto T^{\alpha}$ ($H_s$: the saturation field strength at zero temperature) closer to the HFP approximation $\alpha=1.5$~\cite{oshigawa2000} than that regarding Eq.~(\ref{power_law}) around $H_{c1}$.
{It implies that the strong inter-dimer interactions contribute to a remarkable dispersion in the dimer spin system. This dispersion can be understood by magnon-magnon or hole-hole interaction in a boson system.
Thus we have proved that the dispersion of the particle manifests itself in the asymmetric convex magnetization process, which indicates that a precise measurement of the magnetization can give us microscopic nature of magnons in some particular magnets.}

In summary, we have measured the magnetization process of TlCuCl$_3$ at 2~K and the second critical magnetic field $H_{c2}$ to be 86.1~T.
The magnetization process shows a continuous convex slope and we analyze the critical exponents of magnetization at $H_{c1}$ and $H_{c2}$.
A Monte Carlo calculation based on cubic lattice well reproduces the experimental results and strongly supports the itinerant property of magnons in TlCuCl$_3$.
The {particle-hole symmetry} has been revealed to be broken in TlCuCl$_3$.
We have also found that the degree of asymmetry of magnetization process increases with the inter-dimer interactions by a numerical analysis with the QMC method. Thus, such interactions play an essential role in magnetization of various spin dimer systems.

X.-G.~Z thank Y. Kohama and T. Nomura for fruitful discussion.
X.-G.~Z. was supported by MEXT scholarship.
Y.~Y. was supported by JSPS fellowship.
This work was partially supported by JSPS KAKENHI Grant Nos. JP19J13783 (Y.~Y.).


\sloppy

%


\end{document}